\documentstyle[prd,aps,eqsecnum,preprint,tighten,floats]{revtex}
\begin{document}
\draft
\preprint{UMN-D-97-2}
\title{Nonperturbative Renormalization in Light-Cone 
Quantization\footnote{To appear in the proceedings
of Orbis Scientiae 1997: {\em Twenty-Five Coral Gables
Conferences and their Impact on High Energy Physics 
and Cosmology}, B.N. Kursunoglu, ed., 
(Plenum, New York, 1997).} }

\author{John R. Hiller}
\address{%
Department of Physics, 
University of Minnesota,
Duluth, Minnesota~~55812}
\date{March 20, 1997} 

\maketitle

\begin{abstract}
Two approaches to nonperturbative renormalization are
discussed for theories quantized on the light cone.
One is tailored specifically to a calculation of
the dressed-electron state in quantum electrodynamics,
where an invariant-mass cutoff is used as a regulator
and a Tamm--Dancoff truncation is made to include
no more than two photons.  The other approach is 
based on Pauli--Villars regulators and is applied
to Yukawa theory and a related soluble model.  In both
cases discretized light-cone quantization is used to
obtain a finite matrix problem that can be solved
nonperturbatively.
\end{abstract}

\section{INTRODUCTION}

Light-cone quantization\cite{Dirac} has attracted some interest
as a means to perform nonperturbative analyses of quantum field
theories\cite{Reviews}.  There are good reasons to hope that
this technique will provide the leverage needed to obtain a
qualitative, and perhaps quantitative, connection between
quantum chromodynamics (QCD) and the constituent quark 
model\cite{Wilson}.  Given the complexity of QCD, it is
useful to first study simpler theories such as quantum
electrodynamics (QED) and even models in $1+1$ spacetime
dimensions rather than $3+1$ dimensions.

Bound-state calculations in QCD$_{3+1}$ and QED$_{3+1}$ require 
nonperturbative renormalization.  Most attempts at such calculations 
have used Tamm--Dancoff truncations\cite{TammDancoff} and cutoff-type 
regularization, which require counterterms that depend on Fock 
sector\cite{SectorDependent}.  An example of such a calculation 
is given here for the electron's anomalous moment\cite{Zakopane}.  
We then explore the practicality of Pauli--Villars 
regularization\cite{PauliVillars} as an alternative.  
In particular, we consider a simple heavy-fermion model 
abstracted from the Yukawa model.

We define light-cone coordinates by
\begin{equation}  \label{eq:coordinates}
x^\pm=t\pm z\,,\;\;{\bf x}_\perp=(x,y)\,.
\end{equation}
Momentum variables are similarly constructed as
\begin{equation}  \label{eq:momentum}
p^\pm=E\pm p_z\,,\;\;{\mathbf p}_\perp=(p_x,p_y).
\end{equation}
The dot product is written
\begin{equation}
p\cdot x=\frac{1}{2}(p^+x^-+p^-x^+)
                 -{\bf p}_\perp\cdot{\bf x}_\perp\,.
\end{equation}
The time variable is taken to be $x^+$, and time 
evolution of a system is then determined by the 
conjugate operator ${\cal P}^-$.  The energy $E$ 
is replaced by the light-cone energy $p^-$, and 
the momentum ${\bf p}$ by the light-cone momentum 
$\underline{p}\equiv (p^+,{\bf p}_\perp)$.
The light-cone Hamiltonian is
\begin{equation}  \label{eq:HLC}
H_{\rm LC}={\cal P}^+{\cal P}^- - {\cal P}^2_\perp\,,
\end{equation}
where ${\cal P}^+$ and ${\bf \cal P}_\perp$ are 
momentum operators conjugate to $x^-$ and 
${\bf x}_\perp$.  The eigenvalue problem is
\begin{equation}  \label{eq:EigenProb}
H_{\rm LC}\Psi=M^2\Psi\,,\;\; 
\underline{\cal P}\Psi=\underline{P}\Psi\,,
\end{equation}
where $M$ is the mass of the state.

Some of the advantages of light-cone coordinates are the
following: They admit the largest possible set of 
nondynamical generators.  In particular, boosts are 
kinematical.  For many theories of massive particles, 
the perturbative vacuum is the physical vacuum, because
$p_i^+=\sqrt{p^2+m^2}+p_z>0$ implies that no particle
state can contribute to the $P^+=0$ vacuum.  Thus
there is no need to compute the vacuum state before 
computing massive states.  Also, well-defined Fock-state 
expansions exist, with no disconnected vacuum pieces.

Such expansions are written as 
\begin{equation}  \label{eq:Psi}
\Psi=\sum_n\int [dx]_n\,[d^2k_\perp]_n\,
       \psi_n(x,{\bf k}_\perp)
             |n:xP^+,x{\bf P}_\perp+{\bf k}_\perp\rangle\,,
\end{equation}
with $n$ the number of particles, $i$ ranging between 1 
and $n$, $(P^+,{\bf P}_\perp)$ the total light-cone 
momentum, and
\begin{equation}  \label{eq:dx}
[dx]_n=4\pi\delta(1-\sum_{i=1}^nx_i)
	     \prod_{i=1}^n\frac{dx_i}{4\pi\sqrt{x_i}}\,,\;\;\; 
[d^2k_\perp]_n=4\pi^2\delta(\sum_{i=1}^n{\bf k}_{\perp i})
	       \prod_{i=1}^n\frac{d^2k_{\perp i}}{4\pi^2}\,.
\end{equation}
In the Fock basis 
$\{|n:p_i^+,{\bf p}_{\perp i}\rangle\}$, ${\cal P}^+$ and
${\bf \cal P}_\perp$ are diagonal.
The amplitude $\psi_n$ is interpreted as the wave function 
of the contribution from states with $n$ particles.

A common numerical technique is discretized light-cone 
quantization (DLCQ)\cite{PauliBrodsky}, in which
periodic boundary conditions are assigned to bosons 
and antiperiodic to fermions in a light-cone box 
$-L<x^-<L$, $-L_\perp<x,y<L_\perp$.  Integrals are 
replaced by trapezoidal approximations on a grid:
$p^+\rightarrow\frac{\pi}{L}n$, 
${\bf p}_\perp\rightarrow
     (\frac{\pi}{L_\perp}n_x,\frac{\pi}{L_\perp}n_y)$,
with $n$ even for bosons and odd for fermions.
The limit $L\rightarrow\infty$ can be exchanged for a limit
in terms of the integer {\em resolution}
$K\equiv\frac{L}{\pi}P^+$.
The longitudinal momentum fraction $x=p^+/P^+$ becomes $n/K$.  
$H_{\rm LC}$ is independent of $L$.  

Because the $n_i$ are all positive, DLCQ
automatically limits the number of particles to no more 
than $\sim\!\!K/2$.  The integers $n_x$ and $n_y$ range 
between limits associated with some maximum integer 
$N_\perp$ fixed by $L_\perp$ and a cutoff that 
limits transverse momentum.

To reduce the size of the discrete matrix problem,
a Tamm--Dancoff truncation\cite{TammDancoff} in
the number of particles can be applied.  This has
serious implications for renormalization.
These include severe sector dependence of 
counterterms\cite{SectorDependent}, and, for QED,
violation of the Ward identity.

Regularization via cutoffs typically involves limits
on the invariant mass.  A limit can be placed on the 
total invariant mass of each Fock state
\begin{equation}  \label{eq:cutoff1}
\sum_i \frac{m_i^2+k_{\perp i}^2}{x_i}\leq\Lambda^2
\end{equation}
or on the invariant mass of each particle
\begin{equation}  \label{eq:cutoff2}
\frac{m_i^2+k_{\perp i}^2}{x_i}\leq\Lambda^2\,.
\end{equation}
There can also be a limit on the change in invariant
mass across each matrix element of 
$H_{\rm LC}$\cite{Lepage}
\begin{equation}  \label{eq:cutoff3}
\left|\sum_i^n \frac{m_i^2+k_{\perp i}^2}{x_i}
  -\sum_j^m \frac{m_j^2+k_{\perp j}^2}{x_j}\right|
      \leq\Lambda^2\,.
\end{equation}

\section{THE ANOMALOUS MOMENT}

The anomalous moment $a_e=F_2(0)$ can be computed from a
spin-flip matrix element of the electromagnetic current
\begin{equation}   \label{eq:current}
-\frac{q_1}{2m_e}F_2(q^2)=
  \frac{1}{2P^+}\langle P+q,\uparrow|J^+(0)|P,\downarrow\rangle
\end{equation}
in the standard light-cone frame
$q=(0,q_\perp^2/P^+,{\bf q}_\perp=q_1\hat{x})$.
Brodsky and Drell\cite{BrodskyDrell} have given a useful
reduction of this matrix element to the form
\begin{equation}   \label{eq:aeLC}
a_e=-2m_e\sum_je_j\sum_n\int\,[dx]_n\,[d^2k_\perp]_n\,
    \psi_{n\uparrow}^*(x,{\bf k}_\perp)
    \sum_{i\neq j}x_i\frac{\partial}{\partial k_{1i}}
    \psi_{n\downarrow}(x,{\bf k}_\perp)\,,
\end{equation}
where $e_j$ is the fractional charge of the struck constituent
and $x_i=p_i^+/P^+$.
The wave functions $\psi_n$ satisfy coupled integral equations
obtained from $H_{\rm LC}\Psi=M^2\Psi$.  The QED light-cone 
Hamiltonian has been given by Tang {\em et al.}\cite{Tang}.  
However, the bare masses and couplings must be computed
from sector dependent renormalization conditions.

Consider the case where there are at most two photons
and only one electron.  The Fock-state expansion can be 
written schematically as
\begin{equation} \label{eq:SchematicExpansion}
\Psi=\psi_0|e\rangle+\vec{\psi}_1|e\gamma\rangle
	      +\vec{\psi}_2|e\gamma\gamma\rangle\,.
\end{equation}
Here $\vec{\psi}_1$ and $\vec{\psi}_2$ are column vectors that 
contain the amplitudes for individual Fock states with one and 
two photons, respectively.  The eigenvalue problem becomes a 
coupled set of three integral equations
\begin{eqnarray} \label{eq:IntegralEqns}
m_0^2\psi_0 + {\bf b}_1^\dagger\cdot\vec{\psi}_1
     + {\bf b}_2^\dagger\cdot\vec{\psi}_2  & = & M^2\psi_0\,, 
\nonumber \\
{\bf b}_1\psi_0 + A_{11}\vec{\psi}_1
     + A_{12}\vec{\psi}_2 & = & M^2\vec{\psi}_1\,,   \\
{\bf b}_2\psi_0 + A_{12}^\dagger\vec{\psi}_1
     + A_{22}\vec{\psi}_2 & = & M^2\vec{\psi}_2\,,   \nonumber
\end{eqnarray}
where $m_0$ is the bare electron mass and ${\bf b}_i^\dagger$ 
and $A_{ij}$ are integral operators obtained from 
matrix elements of $H_{\rm LC}$.

The bare electron mass in the one-photon sector is computed
from the one-loop self energy allowed by the two-photon states. 
We then
require that $m_0$ be such that $M^2=m_e^2$ is an eigenvalue.  The
second and third equations can be solved for $\vec{\psi}_1/\psi_0$
and $\vec{\psi}_2/\psi_0$.  Then the first equation yields $m_0$.
Normalization of $\Psi$ fixes the value of $\psi_0$.

The bare coupling for the electron-photon three-point vertex 
depends on the initial and final momenta of the electron and 
on the sectors between which the coupling acts.  The momentum
dependence is present because the amount of momentum available 
constrains the extent to which higher order corrections can 
contribute.  Similarly, the sector dependence makes itself 
felt when the number of additional particles in higher-order 
corrections is restricted.  The coupling is fixed by the 
ratio of the e$\gamma\rightarrow$e transition matrix 
element to the bare vertex at zero photon momentum.

In the present calculation we use a Tamm--Dancoff truncation 
to \{e, e$\gamma$, e$\gamma\gamma$\}, a nonzero photon mass 
$m_\gamma=m_e/10$, and a moderate coupling $\alpha=1/10$.
Some results are given elsewhere\cite{Zakopane}.
When only states with at most one photon and no 
pairs are retained, one can show that $a_e$ 
reduces to
\begin{equation}
a_e=\frac{\alpha m_e^2}{\pi^2}\int\,\frac{dx\,d^2k_\perp}{x}
   \frac{\theta(\Lambda^2-(m_e^2+k_\perp^2)/x
                                -(m_\gamma^2+k_\perp^2)/(1-x))}
          {[m_e^2-(m_e^2+k_\perp^2)/x
                 -(m_\gamma^2+k_\perp^2)/(1-x)]^2}\,,
\end{equation}
which in the limit of $\Lambda\longrightarrow\infty$ 
becomes\cite{BrodskyDrell}
\begin{equation}
a_e=\frac{\alpha}{2\pi}\int_0^1
        \frac{2x^2(1-x)dx}{x^2+(1-x)(m_\gamma/m_e)^2}\,.
\end{equation}
For $m_\gamma=0$, this yields the standard Schwinger 
contribution\cite{Schwinger} of $\alpha/2\pi$.

\section{YUKAWA THEORY AT ONE LOOP}

As an alternative approach to regularization, we consider
Pauli--Villars\cite{PauliVillars} regularization of the 
$3+1$ Yukawa model\cite{Yukawa,BurkardtLangnau}.  
The one-loop fermion self-energy is proportional to 
\begin{equation}
I(\mu^2,M^2)\equiv-\frac{1}{\mu^2}\int
    \frac{dl^+d^2l_\perp}{l^+(q^+-l^+)^2}
       \frac{(q^+)^2{\bf l}_\perp^2+(2q^+-l^+)^2M^2}
           {M^2-D_1}\theta(\Lambda^2-D_1)\,,
\end{equation}
where $\underline{q}$ is the fermion momentum,
$\mu$ is the boson mass, $M$ is the fermion mass, and
\begin{equation}
D_1=\frac{\mu^2+{\bf l}_\perp^2}{l^+/q^+}
               +\frac{M^2+{\bf l}_\perp^2}{(q^+-l^+)/q^+}\,.
\end{equation}
The boson mass $\mu$ sets the energy scale.
When $M^2=0$ we obtain
\begin{equation}
I(\mu^2,0)=\frac{\pi}{\mu^2}\left[\frac{\Lambda^2}{2}-
            \frac{\mu^4}{2\Lambda^2}-
         \mu^2\ln\left(\frac{\Lambda^2}{\mu^2}\right)\right]\,.
\end{equation}
In order to maintain $I(\mu^2,M^2)\propto M^2$,
three Pauli-Villars bosons are needed:\cite{ChangYan}
\begin{equation} \label{eq:Isub}
I_{\rm sub}(\mu^2,M^2,\mu_i^2)=
   I(\mu^2,M^2)+\sum_{i=1}^3 C_i I(\mu_i^2,M^2)\,.
\end{equation}
The $C_i$ are chosen to satisfy
\begin{equation}
1+\sum_{i=1}^3 C_i=0\,, \;\;
\mu^2+\sum_{i=1}^3 C_i\mu_i^2=0\,, \;\;
\sum_{i=1}^3 C_i\mu_i^2\ln(\mu_i^2/\mu^2)=0\,.
\end{equation}

A DLCQ calculation of $I_{\rm sub}$ has been done\cite{BHM},
with values of 20, 22, and 24 for $K$ and 25 through 30 for 
$N_\perp$.  Modification of the trapezoidal rule, with 
introduction of unequal weights, is necessary to obtain 
sufficient accuracy.  Each integral in (\ref{eq:Isub}) was 
separately extrapolated to infinite $K$ and $N_\perp$ via 
fits to either $c_0+a_1/K^3+b_1/N_\perp^2$ or
$c_0+a_1/K^3+a_2/K^4+b_1/N_\perp^2+b_2/N_\perp^3$.
The latter was used for the $\mu_1$ integral.
Extrapolation after subtraction is not as accurate.
The resulting values of $I_{\rm sub}$ were extrapolated
to infinite cutoff by fits to $a+b/\Lambda^2$. 
These fully extrapolated values are given in 
Table~\ref{tab:Linfinity}.

\begin{table}
\caption{Values of the subtracted integral 
$I_{\rm sub}(M^2/\mu^2,\mu_i^2/\mu^2)$ in the
limit of infinite cutoff.  The Pauli--Villars masses
are $\mu_1^2=10\mu^2$, $\mu_2^2=50\mu^2$ and 
$\mu_3^2=100\mu^2$.}
\label{tab:Linfinity}
\begin{tabular}{c|cccc}
$M^2$ & 0 & $0.05\mu^2$ & $0.1\mu^2$ & $0.2\mu^2$ \\
\hline
$I_{\rm sub}$ & -0.064 & 0.70 & 1.37 & 2.70 \\
\end{tabular}
\end{table}

The magnitude of the error in each extrapolated integral was 
found to be $\leq0.02$ when compared to the analytic result 
for $M^2=0$.  This implies an error of $\pm0.04$ in the 
$I_{\rm sub}$ values.  The extrapolation in $\Lambda^2$ 
induces additional uncertainty reflected in the miss
of zero by 0.06 for $M^2=0$.  The values in 
Table~\ref{tab:Linfinity} are consistent with 
$I_{\rm sub}\propto M^2$ to within this amount of error.

The number of Fock states required for Pauli--Villars 
particles is approximately 1.5 times the number for 
physical states.  A listing of counts for two cases 
is given in Table~\ref{tab:FockStates}.  Making $\mu_1$ 
larger does decrease the number of Pauli-Villars states 
but this increases the coefficients $C_i$ and thereby 
amplifies errors in the integrals.  Also, with fewer 
states, the integrals themselves are approximated 
less accurately.

\begin{table}
\caption{Number of Fock states used in two typical cases.}
\label{tab:FockStates}
\begin{tabular}{cccccccc}
            &     &           &  physical  &
     \multicolumn{4}{c}{Pauli-Villars boson states} \\ 
     \cline{5-8}
$\Lambda^2/\mu^2$ & $K$ & $N_\perp$ & boson states & 
     $\mu_1^2=10\mu^2$ & $\mu_2^2=50\mu^2$ & 
                            $\mu_3^2=100\mu^2$ & total\\
\hline 
200  &  20  &  25  &  25975  &  22602  
                      & 11142 & 3305 & 37049 \\
200  &  24  &  30  &  44943  &  39162  
                     &  19293 &  5695  &  64150 \\
\end{tabular}
\end{table}

We could also consider the boson self energy.  To lowest
order there is a fermion loop contribution 
\begin{equation}
\int\frac{dl^+d^2l_\perp}{4LL_\perp^2}
       \frac{q^+(l_\perp^2+M^2)}{l^{+2}(q^+-l^+)^2}
          \frac{\theta\left(\Lambda^2-D_2\right)}{\mu^2-D_2}\,,
\end{equation}
where
\begin{equation}
D_2\equiv q^{+2}(M^2+l_\perp^2)/[l^+(q^+-l^+)]\,,
\end{equation}
and a $\phi^4$ contribution
\begin{equation}
\int\frac{dl^+d^2l_\perp dk^+d^2k_\perp}{q^+l^+k^+(q^+-l^+-k^+)}
 \frac{\theta\left(\Lambda^2-D_4\right)}{\mu^2-D_4}\,,
\end{equation}
where 
\begin{equation}
D_4\equiv\frac{\mu^2+l_\perp^2}{l^+/q^+}
       +\frac{\mu^2+k_\perp^2}{k^+/q^+}
       +\frac{\mu^2+(l_\perp+k_\perp)^2}{(q^+-l^+-k^+)/q^+}\,.
\end{equation}
A Pauli--Villars fermion may be needed.

\section{A HEAVY-FERMION MODEL}

By some severe modifications of the Yukawa 
Hamiltonian\cite{McCartorRobertson} we obtain the 
following model Hamiltonian:
\begin{eqnarray}
H_{\rm LC}^{\rm eff}
   &=&M_0^2\int\frac{dp^+d^2p_\perp}{16\pi^3p^+}
            \sum_\sigma b_{\underline{p}\sigma}^\dagger 
                                b_{\underline{p}\sigma}
     +P^+\int\frac{dq^+d^2q_\perp}{16\pi^3q^+}
         \left[\frac{\mu^2+q_\perp^2}{q^+}
                a_{\underline{q}}^\dagger a_{\underline{q}}
           + \frac{\mu_1^2+q_\perp^2}{q^+}
                a_{1\underline{q}}^\dagger a_{1\underline{q}}
               \right] \nonumber  \\
   &  & +g\int\frac{dp_1^+d^2p_{\perp1}}{\sqrt{16\pi^3p_1^+}}
            \int\frac{dp_2^+d^2p_{\perp2}}{\sqrt{16\pi^3p_2^+}}
              \int\frac{dq^+d^2q_\perp}{16\pi^3q^+}
                \sum_\sigma b_{\underline{p}_1\sigma}^\dagger 
                     b_{\underline{p}_2\sigma}    \\
    &  & \rule{0.5in}{0mm}\times \left[
         a_{\underline{q}}^\dagger
           \delta(\underline{p}_1-\underline{p}_2+\underline{q})
        +a_{\underline{q}}
        \delta(\underline{p}_1-\underline{p}_2-\underline{q}) 
             \right.           \nonumber \\
    &  & \rule{0.75in}{0mm} \left.
       +ia_{1\underline{q}}^\dagger
         \delta(\underline{p}_1-\underline{p}_2+\underline{q})
      +ia_{1\underline{q}}
        \delta(\underline{p}_1-\underline{p}_2-\underline{q}) 
            \right]\,.             \nonumber
\end{eqnarray}
The kinetic energy of the fermion is no longer momentum dependent 
and only a modified no-flip three-point vertex remains as an
interaction.  The fermion then acts as a ``static'' source for
the boson.  We include one Pauli--Villars field, which will prove 
sufficient in this case.  Similar Hamiltonians, without the 
Pauli--Villars field, have been considered in 
equal-time\cite{SchweberGreenberg} and light-cone 
coordinates\cite{GlazekPerry}.  

We write the eigenvector as a Fock-state expansion
\begin{eqnarray}
\Phi_\sigma&=&\sqrt{16\pi^3P^+}\sum_{n,n_1}
                  \int\frac{dp^+d^2p_\perp}{\sqrt{16\pi^3p^+}}
   \prod_{i=1}^n\int\frac{dq_i^+d^2q_{\perp i}}{\sqrt{16\pi^3q_i^+}}
   \prod_{j=1}^{n_1}
        \int\frac{dr_j^+d^2r_{\perp j}}{\sqrt{16\pi^3r_j^+}} \\
   &  & \times \delta(\underline{P}-\underline{p}
              -\sum_i^n\underline{q}_i-\sum_j^{n_1}\underline{r}_j)
       \phi^{(n,n_1)}(\underline{q}_i,\underline{r}_j;\underline{p})
         \frac{1}{\sqrt{n!n_1!}}b_{\underline{p}\sigma}^\dagger
          \prod_i^n a_{\underline{q}_i}^\dagger 
             \prod_j^{n_1} a_{1\underline{r}_j}^\dagger |0\rangle \,,
   \nonumber
\end{eqnarray}
normalized according to
$\Phi_\sigma^{\prime\dagger}\cdot\Phi_\sigma
=16\pi^3P^+\delta(\underline{P}'-\underline{P})$,
which yields
\begin{equation}  \label{eq:NormCondition}
1=\sum_{n,n_1}\prod_i^n\int\,dq_i^+d^2q_{\perp i}
                     \prod_j^{n_1}\int\,dr_j^+d^2r_{\perp j}
    \left|\phi^{(n,n_1)}(\underline{q}_i,\underline{r}_j;
           \underline{P}-\sum_i\underline{q}_i
                       -\sum_j\underline{r}_j)\right|^2\,.
\end{equation}
For $\Phi_\sigma$ to satisfy the Schr\"odinger 
equation (\ref{eq:EigenProb}), the amplitudes must satisfy
\begin{eqnarray}
\left[M^2-M_0^2\rule{0mm}{0.35in}\right.
  &&\left.\rule{0mm}{0.35in}-\sum_i\frac{\mu^2+q_{\perp i}^2}{y_i}
       -\sum_j\frac{\mu_1^2+r_{\perp j}^2}{z_j}
                                      \right]\phi^{(n,n_1)}  \\
& & =g\left\{\sqrt{n+1}\int\frac{dq^+d^2q_\perp}{\sqrt{16\pi^3q^+}}
              \phi^{(n+1,n_1)}(\underline{q}_i,\underline{q},
                             \underline{r}_j,\underline{p})\right.
\nonumber \\
& & \rule{0.75in}{0mm} 
        +\frac{1}{\sqrt{n}}\sum_i\frac{1}{\sqrt{16\pi^3q_i^+}}
          \phi^{(n-1,n_1)}(\underline{q}_1,\ldots,\underline{q}_{i-1},
                          \underline{q}_{i+1},\ldots,\underline{q}_n,
                              \underline{r}_j,\underline{p})
\nonumber \\
& &\rule{0.75in}{0mm}+i\sqrt{n_1+1}
         \int\frac{dr^+d^2r_\perp}{\sqrt{16\pi^3r^+}}
              \phi^{(n,n_1+1)}(\underline{q}_i,\underline{r}_j,
                                     \underline{r},\underline{p})
\nonumber \\
& & \rule{0.75in}{0mm} 
      +\left.\frac{i}{\sqrt{n}}\sum_j\frac{1}{\sqrt{16\pi^3r_j^+}}
       \phi^{(n,n_1-1)}(\underline{q}_i,
                        \underline{r}_1,\ldots,\underline{r}_{j-1},
                       \underline{r}_{j+1},\ldots,\underline{r}_{n_1},
                                \underline{p}) \right\}\,.   \nonumber
\end{eqnarray}
The structure of this coupled set of integral equations is deliberately
identical in basic form to the equations considered by Greenberg and 
Schweber\cite{SchweberGreenberg}.  Therefore, we transcribe their 
{\em ansatz} for a solution to light-cone form
\begin{equation}
\phi^{(n,n_1)}=\sqrt{Z}\frac{(-g)^n(-ig)^{n_1}}{\sqrt{n!n_1!}}
       \prod_i\frac{q_i^+}{\sqrt{16\pi^3q_i^+}(\mu^2+q_{\perp i}^2)}
       \prod_j\frac{r_j^+}{\sqrt{16\pi^3r_j^+}(\mu_1^2+r_{\perp j}^2)}\,.
\end{equation}
This does work as a solution if $M_0^2(\mu_1)$ is chosen to satisfy
\begin{equation}
M^2-M_0^2=-\frac{g^2}{16\pi^3}
           \left\{\int\frac{dyd^2q_\perp}{\mu^2+q_\perp^2}
              -\int\frac{dzd^2r_\perp}{\mu_1^2+r_\perp^2}\right\}\,.
\end{equation}
From the normalization condition (\ref{eq:NormCondition}) we obtain
\begin{equation}
\frac{1}{Z}=\exp\left\{\frac{g^2}{16\pi^3}
         \left[\int\frac{ydyd^2q_\perp}{(\mu^2+q_\perp^2)^2}
            +\int\frac{zdzd^2r_\perp}{(\mu_1^2+r_\perp^2)^2}
                                              \right]\right\}\,.
\end{equation}
The bare mass and wave function renormalization are thus determined
as functions of the Pauli--Villars mass.

To fix the coupling we could use the slope of the fermion no-flip form
factor, which is related to the transverse size of the dressed fermion.
The form factor is most easily evaluated from\cite{BrodskyDrell} 
\begin{eqnarray}
F(Q^2)&=&\frac{1}{2P^+}
          \langle P+p_\gamma\uparrow |J^+(0)|P\uparrow\rangle \\
      &=&\sum_j e_j\int 16\pi^3\delta(1-\sum_i x_i)
                   \delta(\sum_i {\bf k}_{\perp i})
      \prod_i \frac{dx_id^2p_{\perp i}}{16\pi^3} \nonumber \\
      & & \rule{0.75in}{0mm} 
            \times \psi_{P+p_\gamma\uparrow}^*(x_i,{\bf p}'_{\perp i})
              \psi_{P\uparrow}(x_i,{\bf p}_{\perp i})\,,  \nonumber
\end{eqnarray}
where the matrix element has been evaluated in the frame with
\begin{equation}
P=(P^+,P^-=\frac{M^2}{P^+},{\bf 0}_\perp)\,,\;\;
p_\gamma=(0,p_\gamma^-=2p_\gamma\cdot P/P^+,
                               {\bf p}_{\gamma\perp})\,,\;\;
Q^2\equiv p_{\gamma\perp}^2\,,
\end{equation}
$e_j$ is the charge of the jth constituent, and
\begin{equation}
{\bf p}'_{\perp i}=
   \left\{\begin{array}{cc} 
     {\bf p}_{\perp i}-x_i{\bf p}_{\gamma\perp} & i\neq j \\
     {\bf p}_{\perp i}+(1-x_i){\bf p}_{\gamma\perp} & i=j\,. 
                                           \end{array}\right.
\end{equation}
A sum over Fock states is understood.

When the fermion is assigned a charge of 1, and the bosons 
remain neutral, the analytic solution for the amplitudes 
yields
\begin{equation}
F(Q^2)=Z\exp\left\{g^2\int\frac{dy d^2q_\perp}{16\pi^3}
                \frac{\sqrt{y}}{\mu^2+q_\perp^{\prime 2}}
                 \frac{\sqrt{y}}{\mu^2+q_\perp^2}
                                  +\mbox{P-V term}\right\}\,,
\end{equation}
with  
\begin{equation}
{\bf q}'_\perp={\bf q}_\perp-y{\bf p}_{\gamma\perp}\,.
\end{equation}
From this we find
\begin{equation}
F'(0)=-g^2\int\frac{dyd^2q_\perp}{16\pi^3}
       \frac{y^3}{(\mu^2+q_\perp^2)^3}
       \left[\frac{2\mu^2}{\mu^2+q_\perp^2}-1\right]
                 +\mbox{P-V term}\,.
\end{equation}
Numerically, the slope is computed from a 
finite-difference approximation to
\begin{eqnarray}
F'(0)&=&\sum_{n,n_1}\prod_i^n\int\,dq_i^+d^2q_{\perp i}
              \prod_j^{n_1}\int\,dr_j^+d^2r_{\perp j} \\
    & & \times 
       \left[\left(\sum_i \frac{y_i^2}{4}\nabla_{\perp i}^2+
         \sum_j \frac{z_j^2}{4}\nabla_{\perp j}^2\right)
          \phi^{(n,n_1)}(\underline{q}_i,\underline{r}_j;
       \underline{P}-\sum_i\underline{q}_i-\sum_j\underline{r}_j)
                                          \right]^*
    \nonumber \\
    & & \rule{1in}{0mm} \times
        \phi^{(n,n_1)}(\underline{q}_i,\underline{r}_j;
        \underline{P}-\sum_i\underline{q}_i-\sum_j\underline{r}_j)\,.
    \nonumber
\end{eqnarray}

With the bare parameters determined, we ``predict'' a value for 
$\langle :\!\!\phi^2(0)\!\!:\rangle$.  For the analytic
solution, this expectation value reduces to
\begin{equation}
\langle :\!\!\phi^2(0)\!\!:\rangle=
       \frac{g^2}{8\pi^2\mu^2}\left[1-\frac{\mu^2}{\Lambda^2}
              -\frac{\mu^2}{\Lambda^2}\ln\frac{\mu^2}{\Lambda^2}
                                                      \right]\,.
\end{equation}
From a numerical solution it can be computed from a sum
similar to the normalization sum
\begin{eqnarray}
\langle :\!\!\phi^2(0)\!\!:\rangle
        &=&\sum_{n=1,n_1=0}\prod_i^n\int\,dq_i^+d^2q_{\perp i}
                     \prod_j^{n_1}\int\,dr_j^+d^2r_{\perp j}
                     \left(\sum_{k=1}^n \frac{2}{q_k^+/P^+}
                                                 \right) \\
    & & \rule{0.75in}{0mm}
       \times \left|\phi^{(n,n_1)}(\underline{q}_i,\underline{r}_j;
       \underline{P}-\sum_i\underline{q}_i-\sum_j\underline{r}_j)
                                                  \right|^2\,.
    \nonumber
\end{eqnarray}

\section{SUMMARY}

For the anomalous moment calculation there remain 
several hurdles.  The Tamm--Dancoff truncation results 
in logarithmically divergent four-point graphs.  To 
deal with these will probably require use of scattering 
processes, such as Compton scattering\cite{MustakiPinsky}, 
to obtain renormalization conditions.  Verification of 
the removal of all logarithms and restoration of symmetries 
can then be undertaken.  Also neglected up to this point 
have been zero modes, photon modes of zero longitudinal
momentum\cite{ZeroModes}.  How they might be included has 
been indicated by Kalloniatis and 
Robertson\cite{KalloniatisRobertson}.

Additional physics could be included in the calculation by
introducing an effective interaction from Z graphs or even
putting eee$^+$ states in the basis.  In the latter case,
photon mass renormalization must be done.

In the Yukawa-model calculations we have learned that 
Pauli--Villars Fock states increase the basis size by 
only 150\%, which may not be prohibitive.  To perform 
such calculations accurately with a minimal basis size, 
improvement of ordinary DLCQ, by inclusion of weighting
factors, is critical.

We have found a simple $3+1$ model, related to Yukawa theory,
which can be solved analytically.  Here we will attempt
a nonperturbative numerical solution to further test the
use of Pauli--Villars regularization in DLCQ.  If successful,
we can begin to increase the complexity of the model,
eventually reaching the full Yukawa theory.

\section*{ACKNOWLEDGMENTS}

This work was supported in part by the Minnesota 
Supercomputer Institute through grants of computing 
time.  It was done in collaboration with S.J. Brodsky 
and G. McCartor.

\end{document}